\documentclass[twocolumn,aps,prl]{revtex4}
\usepackage{docs}
\usepackage{graphicx}

\newcommand{\infsum}[1]{\displaystyle\sum_{i = #1}^{\infty}}
\newcommand{\NI}{N_{\mathrm{I}}}
\newcommand{\nion}{n_{\mathrm{I}}}
\newcommand{\nel}{n_{\mathrm{e}}}
\newcommand{\kB}{k_{\mathrm{B}}}
\newcommand{\Ftot}{F_{\mathrm{total}}}
\newcommand{\Ntot}{N_{\mathrm{total}}}
\newcommand{\Vtot}{V_{\mathrm{total}}}
\newcommand{\Zavg}{\langle Z \rangle}
\newcommand{\Smix}{\Delta S_{\mathrm{mixing}}}
\newcommand{\df}{\Delta f_i}
\newcommand{\istar}{i_{*}}
\newcommand{\dfstar}{\Delta f_{i_{*}}}
\newcommand{\omegap}{\omega_{\mathrm{p}}}
\newcommand{\Gammam}{\Gamma_{\mathrm{m}}}
\newcommand{\JMD}{J_{\mathrm{MD}}}

\begin{document}

\title{Classical Nucleation Theory of the One-Component Plasma}

\author{Randall L.\ Cooper$^{1,2}$ and Lars Bildsten$^{2,3}$}
\affiliation{$^1$Harvard-Smithsonian Center for Astrophysics, 60 Garden Street, Cambridge, MA 02138\\   $^2$Kavli Institute for Theoretical Physics, Kohn Hall, University of California, Santa Barbara, CA 93106\\   $^3$Department of Physics, Broida Hall, University of California, Santa Barbara, CA 93106}

\begin{abstract}

We investigate the crystallization rate of a one-component plasma
(OCP) in the context of classical nucleation theory.  From our
derivation of the free energy of an arbitrary distribution of solid
clusters embedded in a liquid phase, we derive the steady-state
nucleation rate of an OCP as a function of the Coulomb coupling
parameter $\Gamma$. Our result for the rate is in accord with recent
molecular dynamics simulations, but it is greater than that of previous
analytical estimates by many orders of magnitude.  Further molecular
dynamics simulations of the nucleation rate of a supercooled liquid
OCP for several  values of $\Gamma$ would  clarify the
physics of this process.

\end{abstract}

\maketitle

\section{Introduction}

The one-component plasma (OCP) consists of a single species of $N$
charged ions of mass $m$ in a volume $V$ in a uniform neutralizing
electron background. This idealized system is a model for
astrophysical settings such as neutron star crusts and white dwarf
interiors.  The dimensionless coupling parameter
\begin{equation}
\Gamma = \frac{(Ze)^{2}}{a \kB T}
\end{equation}
characterizes the state of the OCP, where $Ze$ is the ion's charge, $a
= (3/4 \pi \nion)^{1/3}$ is the mean distance between ions at number
density $\nion=N/V$, and $T$ is the temperature.  We consider the ions
to be classical with a deBroglie wavelength $\lambda = (h^2/2\pi m \kB
T)^{1/2} \ll a$. For $\Gamma \ll 1$, the Coulomb coupling is weak and
the ions behave like an ideal Maxwell-Boltzmann gas.  For $\Gamma \gg
1$, the Coulomb coupling is large and the system is in the liquid
phase until $\Gamma > \Gammam \approx 175$ \citep{PC00,DSBY01}, when
the ions undergo a first-order phase transition into a periodic
lattice with a body-centered cubic configuration.

The Helmholtz free energy of the OCP in the liquid phase is written as
$F=F^{\rm id}+F^{\rm ex}$, where the ideal part $F^{\rm id}=N\kB
T\left[\ln(\nion \lambda^3)-1\right]$, and interactions are
incorporated into the 'excess' part $F^{\rm ex}=-\kB T \log(Z_N/V^N)$,
where $Z_N$ is the configuration integral \citep{HM06}. In the
non-interacting limit, $Z_N\rightarrow V^N$, and $F^{\rm
ex}\rightarrow 0$. Simulations of the liquid OCP have generated
accurate fitting formulas for the reduced excess free energy $f^{\rm
ex}(\Gamma) =F^{\rm ex}/N\kB T$ \citep[e.g.,][]{DS03}, and
calculations of the solid state free energy \citep[e.g.,][]{FH93}
allow for the determination of $\Gammam$.

Crystallization of an OCP occurs via nucleation, the process by which
small crystals form via localized fluctuations in the liquid and grow
via the accretion of surrounding ions \citep[see e.g.,][]{K91}.
Empirically, most pure liquids can be supercooled (in our case $\Gamma
> \Gammam$) without solidifying, which implies that a kinetic barrier
to nucleation exists \citep[e.g.,][]{T52}. Monte Carlo
\citep{O92,DSY93} and molecular dynamics \citep{D06b} simulations
demonstrate that such a kinetic barrier exists in an OCP as well.
Localized fluctuations of sufficient amplitude can generate crystals
large enough to overcome the kinetic barrier, and hence they are
thermodynamically stable.  Such crystals subsequently grow and
facilitate the phase transition.
 
The rate at which astrophysical solids grow can affect the degree of
chemical phase separation during the solidification of both white
dwarf interiors \citep[][and references therein]{HL03} and neutron
star crusts \citep{HBB07}, the existence of an amorphous glassy state
in white dwarf interiors \citep{II83,TI87}, and the prevalence of
defects and impurities in neutron star crusts \citep{D95,DL98}.
Furthermore, the time at which the latent heat of crystallization is
released within a white dwarf affects its observed cooling rate
\citep[e.g.,][and references therein]{FBB01}.  Unfortunately, the
nucleation rate of an OCP is not well understood. Several authors
\citep{IIMI83,IT86,TI87,D95} assumed that classical nucleation theory
\citep{K91} applied to the OCP, but the accuracy of their assumptions
was untested. Daligault \citep{D06b} was the first to directly study
nucleation of an OCP using molecular dynamics simulations.  Although
his results were qualitatively consistent with classical nucleation
theory, it was unclear whether or not the results agreed
quantitatively.  In this paper, we reexamine the nucleation of an OCP
in the context of classical nucleation theory, but with a different
approach to the statistical physics of an ensemble of solid clusters
embedded in a liquid phase. We find a steady-state nucleation rate as
a function of $\Gamma$ in quantitative accord with those of
\citep{D06b}.

We begin in Sec.\ \ref{equilibriumdist} by reviewing what is known
about the excess free energies for multi-component plasmas and then
deriving the minimum reversible work of cluster formation and the
equilibrium distribution of cluster sizes. We use these results to
deduce the nucleation rate in Sec.\ \ref{steadystaterate}, and we
compare this rate, along with that used in previous work, to the
results of \citep{D06b} in Sec.\ \ref{comparison}.  We close in Sec.\
\ref{discussion} by discussing our results and suggesting future
numerical experiments and analytic improvements to this work.
 
\section{Minimum Reversible Work of Cluster Formation}\label{equilibriumdist}

 We consider a classical OCP consisting of $\NI$
ions of mass $m$ in a uniform neutralizing background of electrons at
$\nel = Z \nion$. Motivated by the typical astrophysical context, we
also assume that degenerate electrons supply all of the pressure so
that $\nel$ is constant during any phase change and set by charge
neutrality.  However, before discussing how we treat solid clusters
embedded in a liquid OCP, it is important to clarify what is known
about mixtures of ions of different charges.

Hansen {\it et al.~}\citep{HTV77} first showed that the excess
Helmholtz free energy in a mixture of ions of charges $Z_i$ and number
$N_i$ in a volume $V$ is simply given by the sums of the excess free
energies of each separate species at the same electron density,
\begin{equation} 
F^{\rm ex}=\kB T \displaystyle\sum_{i} N_i f^{\rm ex}(\Gamma_i).
\end{equation} 
Here $\Gamma_i=Z_i^{5/3}\Gamma_e$ and $\Gamma_e=e^2/a_e \kB T$, where
$a_e= (3/4 \pi \nel)^{1/3}$ is the mean distance between electrons, 
and all $\Gamma_i > 1$.
Recent simulations \citep{DS03} have shown that the deviations from
this simple linear mixing rule are less than $0.05\%$. Hence, to an
excellent approximation, the free energy of a mixture of liquid ions
is simply
\begin{equation} 
{F_{\rm total}\over \kB T} = \displaystyle\sum_{i} N_i \left[\ln
\left({N_i\lambda_i^3\over V}\right)-1\right] +\displaystyle\sum_{i}
N_i f^{\rm ex}(\Gamma_i).
\end{equation} 
Later authors chose to write this equation differently by using the
fits, $f_l$, of the total Helmholtz free energies of each pure state
and then mixing them.  This results in the more often seen equation 
\begin{equation} {F_{\rm total}\over \kB T} =\displaystyle\sum_{i} N_i 
f_l(\Gamma_e Z_i^{5/3}) + \displaystyle\sum_{i} N_i \ln
\left({Z_i N_i \over \Zavg \Ntot} \right),
\end{equation} 
where $\Zavg$ is the average charge.  The last term is minus the
entropy of mixing \citep[see e.g.,][and the Appendix]{HTV77,BHJ79},
although it originates simply from charge conservation at a fixed
$\nel$.  This term will appear in our derivation of the minimum
reversible work of cluster formation.

This knowledge of mixing at fixed $\nel$ motivates our
re-derivation of the minimum reversible work of cluster formation, and
thereby the equilibrium cluster distribution, from the total Helmholtz
free energy $\Ftot$ of an arbitrary distribution of clusters.
Following Frenkel \citep{F39} (see also \cite{A74,V06}), we consider
clusters that contain different numbers of ions to be distinct
species, and we consider all clusters that contain the same number of
ions to be indistinguishable.  The latter is an approximation because
such clusters could have different morphologies.  Thus we assume that,
for each cluster size, there exists a unique configuration of ions for
which the total surface free energy is a minimum, that this
configuration is spherical, and that all clusters of this size conform
to this configuration.

The total Helmholtz free energy of the system is
\begin{eqnarray}\label{Ftotal}
\frac{\Ftot}{\kB T} &=& N_1 f_l + \infsum{2} N_i (i f_s + i^{2/3}
\alpha) \nonumber \\ &&+ \infsum{1} N_i \left [\ln
\left(\frac{N_i}{\Ntot}\right) + \ln \left (\frac{Z_i}{\Zavg} \right )
\right ],
\end{eqnarray}
where $f_l \kB T$ and $f_s \kB T$ are the free energies of one ion in
the pure liquid and pure solid state, respectively, $\alpha \kB T$ is
proportional to the surface free energy of an ion at the boundary of a
cluster, $N_i$ and $Z_i = iZ$ are the number and charge, respectively,
of clusters (including liquid monomers) that consist of $i$ ions,
\begin{equation}
\Ntot = \displaystyle\sum_{i=1}^{\infty} N_i
\end{equation} 
is the total number of clusters (including monomers), and
\begin{equation}
\Zavg = \displaystyle\sum_{i=1}^{\infty} Z_i \frac{N_i}{\Ntot}
\end{equation} 
is the average charge of the clusters.  Equation (\ref{Ftotal}) is
subject to the constraint
\begin{equation}\label{totalions}
\NI = \displaystyle\sum_{i=1}^{\infty} i N_i.
\end{equation}

From equation (\ref{Ftotal}), the chemical potentials of monomers and
clusters are
\begin{eqnarray}
\frac{\mu_1}{\kB T} &=& f_l + \ln \left(\frac{Z_1 N_1}{\Zavg
\Ntot}\right) + 1 - \frac{Z_1}{\Zavg},
\label{chempot1}\\
\frac{\mu_i}{\kB T} &=& i f_s + i^{2/3} \alpha + \ln \left(\frac{Z_i
N_i}{\Zavg \Ntot}\right) + 1 - \frac{Z_i}{\Zavg},
\label{chempoti} 
\end{eqnarray}
respectively.  Equation (\ref{chempot1}), the chemical
potential of a liquid monomer, is identical to that of
\citep{Ietal91}.  According to classical nucleation theory, clusters
grow or shrink by the addition or subtraction of single monomers.
Thus in chemical equilibrium $\mu_i = \mu_{i-1} + \mu_1$, and so
\begin{equation}\label{chemeq}
\mu_i = i \mu_1
\end{equation}
by iteration.  From equations (\ref{chempot1}), (\ref{chempoti}), and
(\ref{chemeq}), and noting that $\Zavg \Ntot = Z \NI$, the equilibrium
distribution of clusters is
\begin{equation}\label{clusterdist}
\frac{i N_i}{\NI} = \left (\frac{N_1}{\NI} \right )^{i} \exp \left
(-[i(f_s-f_l-1) + i^{2/3} \alpha + 1] \right ),
\end{equation}
which is in the form of the law of mass action.  The
monomer concentration, and hence the concentration of all clusters, is found by
invoking the constraint of equation (\ref{totalions}).  Alternatively, 
equation (\ref{clusterdist}) can be written in terms of the Boltzmann
distribution $P(i) = \exp (-\df)$, where $P(i) = i N_i/\NI$ is the
probability that an ion picked at random is a member of a cluster of
size $i$ and
\begin{equation}\label{reversework}
\df = \left \{ \begin{array}{ll} - \ln (N_1/\NI), & i = 1,\\
i[f_s-f_l-1- \ln (N_1/\NI)] + i^{2/3} \alpha + 1, & i \geq 2
\end{array} \right.
\end{equation}
is the minimum reversible work divided by $\kB T$ of forming a cluster
of size $i$. Demanding that $P(i)$ is normalized then yields 
$N_1/\NI$. 

\section{Steady-State Nucleation Rate}\label{steadystaterate}

We follow the treatment of classical nucleation theory by Kelton
\citep{K91} and derive the steady-state cluster nucleation rate.  We
consider an OCP at fixed $\Gamma$ that is entirely in the liquid state
at time $t = 0$ and calculate the rate of cluster formation for $t >
0$, when $N_1 \approx \NI$, and so
\begin{equation}\label{reversework2}
\df = i ( f_s-f_l-1) + i^{2/3} \alpha + 1,
\end{equation}
from equation (\ref{reversework}).  We use the fit of \citep{PC00}
for $f_l(\Gamma)$, where we take the coefficients derived from 
the simulation data of \citep{DS99}, and we use the fit of \citep{FH93} for 
$f_s(\Gamma)$.  For
sufficiently large $\Gamma$, $(f_s-f_l-1)<0$.  The fit of $f_s$ given
by equations (10) and (15) of \citep{FH93} is valid only for $170 \leq \Gamma
\leq 2000$, and the above inequality is easily satisfied for the entire
range of $\Gamma$ in which the fits are applicable.  The minimum
$\Gamma$ above which $(f_s-f_l-1)<0$ is unknown.  Since $\alpha > 0$,
there exists a critical size
\begin{equation}\label{istar}
\istar = \left ( \frac{2\alpha/3}{1+f_l-f_s} \right )^{3}
\end{equation}
such that the reversible work of cluster formation, $\df$, is a
maximum.  Cluster growth is energetically favorable for clusters of
size $i > \istar$ and unfavorable for clusters of size $i < \istar$.
Equation (\ref{reversework2}) can now be written in the convenient
form
\begin{equation}\label{fstar}
\df = \frac{(2\alpha/3)^{3}}{2 (1+f_l-f_s)^{2}} \left [ 3 \left
(\frac{i}{\istar} \right )^{2/3} - 2 \left (\frac{i}{\istar} \right )
\right ] + 1.
\end{equation}

In classical nucleation theory, crystallization occurs via the
formation and subsequent growth of critical clusters of size $\istar$.
Clusters of size $i < \istar$ are formed via equilibrium fluctuations
and are presumed to be transient.  Equation (\ref{clusterdist}) gives
their distribution.  Clusters of size $i > \istar$ are presumed to be
stable.  Such clusters only grow with time.  The formation of stable
clusters is therefore a two-step process: a stable cluster forms when
(1) equilibrium fluctuations in the liquid phase generate a transient
cluster of size $\istar$, and (2) that cluster accretes an additional
monomer.  Therefore, the steady-state nucleation rate $J$, defined to
be the number of solid clusters formed per second, is roughly equal to
the number of critical clusters of size $\istar$ times the rate at
which a monomer attaches to a critical cluster.  Kelton \citep{K91}
performs a more thorough derivation of the nucleation rate and finds
\begin{equation}\label{nucrateeq}
J (\NI,\Gamma) = \frac{24 D \istar^{2/3}}{a^2} \NI \left
(\frac{\dfstar}{3 \pi \istar^{2}} \right )^{1/2} \exp(-\dfstar),
\end{equation}
where $D$ is the diffusion coefficient and $(\dfstar/3 \pi
\istar^{2})^{1/2}$ is the Zeldovich factor.  We use equation
(\ref{nucrateeq}) for the steady-state nucleation rate of an OCP.  For
$D$, we use the classical OCP diffusion coefficient \citep{HMP75}
\begin{equation}\label{diffusioneq}
D = 2.95 \omegap a^{2} \Gamma^{-1.34},
\end{equation}
where $\omegap = [4 \pi \nion (Ze)^{2}/m]^{1/2}$ is the ion plasma
frequency.  Recent calculations have verified the accuracy of this
expression for $\Gamma \lesssim \Gammam$ \citep{DKG02,DM05} as well as
for $\Gamma \gtrsim \Gammam$ \citep{D06a}.  For the normalized surface
free energy $\alpha$, we follow \citep{D95} and \citep{IIMI83} and set
\begin{equation}\label{alphaeq}
\alpha = 4 \pi \left(\frac{3}{4 \pi} \right )^{2/3} \beta \lambda
\frac{\kB T_\mathrm{m}}{\kB T} = (36 \pi)^{1/3} \beta \lambda \left (
\frac{\Gamma}{\Gammam} \right ),
\end{equation}
where $\beta \approx 0.5$ is the ratio of the interfacial energy and
heat of fusion empirically derived from liquid metal experiments
\citep[e.g.,][]{T50,M95}, and $\lambda \kB T_{\mathrm{m}}$ is the
enthalpy of melting per ion.  For this work, we set $\lambda = 0.77$
\citep{FH93,Setal00}.

\section{Comparison to Previous Work}\label{comparison}

Several authors have investigated the homogeneous nucleation rate of
an OCP \citep{IIMI83,IT86,TI87,D95}.  However, in deriving the minimum
reversible work of cluster formation $\df$, they assumed
that the entropy of mixing is simply that of an ideal gas, i.e.,
the term $\ln (Z_{i}/\Zavg)$ in equation (\ref{Ftotal}) was omitted.
Consequently, they find $\df = i ( f_s-f_l) + i^{2/3} \alpha$, and
 equations (\ref{istar}) and (\ref{fstar}) become
\begin{eqnarray}
\istar &=& \left ( \frac{2\alpha/3}{f_l-f_s} \right
)^{3},\label{istarold}\\ \df &=& \frac{(2\alpha/3)^{3}}{2
(f_l-f_s)^{2}} \left [ 3 \left (\frac{i}{\istar} \right )^{2/3} - 2
\left (\frac{i}{\istar} \right ) \right ].\label{fstarold}
\end{eqnarray}
We demonstrate below that previous calculations 
result in a nucleation rate that is orders of magnitude
lower that the rates deduced from simulations, whereas our model
predicts a rate closer to that found numerically. 
The steady-state nucleation rate per ion as a function of
$\Gamma$ for both models is shown in Figure~1.

Presently, crystallization of an OCP is studied only via numerical
simulations.  Although there are many calculations of the free energy
difference of the pure liquid and pure solid states as a function of
$\Gamma$, few have investigated the onset of crystallization.  Ogata
\citep{O92} and DeWitt {\it et al.~}\citep{DSY93} studied nucleation
in an OCP using Monte Carlo simulations.  Unfortunately, time is not a
variable in Monte Carlo simulations, so that the rate of nucleation is
difficult to calculate \citep[although see][]{OI88}.

Recently, Daligault \citep{D06b} performed molecular dynamics
simulations to investigate the kinetics of nucleation in a supercooled
OCP.  Using $\NI = 4394$ ions, he calculated the total number of solid
nuclei as a function of time for $\Gamma = 400$ and $300$.  In the
$\Gamma = 400$ calculation (see Figure 4 of \citep{D06b}), following a
short period of transient nucleation, there is a significant period of
time during which steady-state nucleation takes place, as evidenced by
the nearly constant slope in the plot of the number of clusters as a
function of time.  In the $\Gamma = 300$ calculation (see Figure 8 of
\citep{D06b}), there is no obvious period of time during which
steady-state nucleation occurs.  To be consistent with our estimate
from the previous calculation, we estimate $\JMD$, the steady-state
nucleation rate deduced from molecular dynamics simulations, from the
period of time during which the nucleation rate is nearly constant,
which occurs just before nucleation saturates.  In Table~I, we list
the inferred values of $\JMD$ as well as the nucleation rates
calculated using both equations (\ref{istar}-\ref{fstar}) of our model
and equations (\ref{istarold}-\ref{fstarold}) from previous models.
Our model predicts nucleation rates that are in reasonable agreement
with the results of Daligault \citep{D06b}, whereas the predictions of
previous models are far too low. 

\begin{figure}\label{nucrateplot1}
\includegraphics[width=\columnwidth]{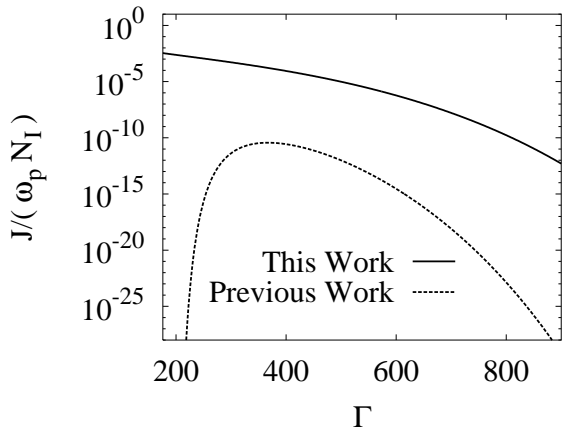}
\caption{Steady-state nucleation rate per ion as a function of $\Gamma$.  
``This Work" is the rate calculated from our model, and ``Previous Work" is 
the rate calculated using the models of previous authors.  Our calculations 
predict that the nucleation rate of a one-component plasma is $> 10^{5}$ 
times higher than previously assumed for all values of $\Gamma$.}
\end{figure}

\begin{table}
\label{nucratetable}
\caption{Steady-state nucleation rates and critical cluster sizes for
$\NI = 4394$: ``$\JMD$" is the inferred nucleation rate from the
simulations of \citep{D06b}, ``(this work)" denotes a quantity
calculated from our model, and ``(previous)" denotes a quantity
calculated using the model of previous authors.}
\begin{ruledtabular}
\begin{tabular}{c c c c c c}
$\Gamma$ & $\JMD/\omegap$ & $J/\omegap$  & $J/\omegap$ & $\istar$ & $\istar$\\
                     &                                 & (this work)       & (previous)& (this work) & (previous)\\
\hline
$400$ & $\approx 0.2$ & $0.4$ & $1 \times 10^{-7}$ & $4$ & $51$\\
$300$ & $\sim 0.1$       & $2$     & $2 \times 10^{-8}$ & $3$ & $91$
\end{tabular}
\end{ruledtabular}
\end{table}

\section{Discussion}\label{discussion}

 Through an improved treatment of the statistical physics of the
distribution of solid clusters in the liquid phase, we have developed
a model of steady-state nucleation in an OCP, and the results of our
model are in accord with those of the time-dependent molecular
dynamics simulations of \citep{D06b}. Our results suggest that
crystallization in an OCP occurs at a rate that is more than five
orders of magnitude higher than previously assumed.

Nucleation in an OCP differs from nucleation in many other media
because of two unique features of an OCP: (1) The degenerate electron
background, which supplies the pressure, is unaffected during the
phase transition.  Consequently, both pressure and volume remain
constant.  Although the complete non-interaction of the electrons is
an approximation of the OCP model, the role electrons play in more
realistic models is negligible \citep{PC00}.  (2) The binding of ions
of like charge in a crystal is weak.  This is reflected in the
difference of the liquid and solid free energies.  Indeed, this
difference does not exceed $\kB T$ (i.e.~$f_l-f_s$ does not exceed
$1$) until $\Gamma/\Gammam = T_{\mathrm{m}}/T \gtrsim 2.8$.  The
combination of these two features has interesting implications for the
crystallization of an OCP.  In particular, nucleation in an OCP is
driven to a significant degree simply by the increase in entropy
resulting from cluster formation: the change in the total energy of a
supercooled OCP when a cluster forms is small relative to the increase
in the entropy.  In fact, equation (\ref{reversework2}) implies that a
population of solid clusters will exist at a given time for a range of
$\Gamma < \Gammam$ because the entropy of an ensemble of clusters
embedded in a liquid OCP is higher than that of the pure liquid phase.
This may be related to the caging effect observed in molecular
dynamics simulations of an OCP in the liquid phase
\citep{DKG02,DHK03}.  Furthermore, we note that a reanalysis of the
entropy of mixing term and thereby the classical nucleation rate may
be needed for other supercooled liquids.

There are a few important issues that we have not addressed.  The
surface free energy of a solid cluster in a Coulomb liquid is not
known.  The expression for the surface free energy contribution used
in this work is only our best estimate, but we cannot be confident
that the numerical coefficient in equation (\ref{alphaeq}) is
accurate.  This issue requires further investigation via 
molecular dynamics simulations.  Some of the
assumptions we made in Sec.\ \ref{equilibriumdist} may not be valid.
We assume that solid clusters grow or shrink by the addition or
subtraction of liquid monomers, but the simulations of Daligault
\citep{D06b} imply both that two solid clusters can fuse together and
that a solid cluster can fission into two or more clusters.
Furthermore, we assume that all clusters of a given size have the same
morphology and are thus indistinguishable, but the simulations of
Daligault \citep{D06b} show that two clusters of the same size can in
fact have different morphologies.  We currently do not know what
effects, if any, a violation of either of these assumptions would have
on our results.  

Further molecular dynamics simulations are imperative to understand
nucleation in an OCP.  It is evident from equations
(\ref{nucrateeq}-\ref{alphaeq}) that the steady-state nucleation rate
may be written as
\begin{equation}
J = \NI \omegap F(\Gamma),
\end{equation}
where $F(\Gamma)$ is a function only of $\Gamma$.
Numerically confirming that the nucleation rate scales linearly with
$\NI$ and $\omegap$ should be simple. One could then deduce
$F(\Gamma)$, and hence $J$, by measuring the nucleation rate for
several different values of $\Gamma$.  Calculations for $\Gammam
\lesssim \Gamma \lesssim 210$ would be particularly useful. Figure 1
shows that there is an enormous difference between our results and
those of previous authors: our model predicts that some clusters
should form rather quickly, whereas other models predict that clusters
essentially never form.

\begin{acknowledgments}
It is a pleasure to thank Niayesh Afshordi, Philip Chang, Burkhard
Militzer, and Ramesh Narayan for helpful discussions, and 
Charles Horowitz and Hugh DeWitt for their critiques of 
earlier drafts of this manuscript. This research
was supported  by the National Science Foundation under grants
PHY 05-51164 and AST 07-07633. 
\end{acknowledgments}

\begin{appendix}*
\section{Entropy of Mixing in a One-Component Plasma}

We follow the treatment of \citep{P96} and derive the ideal entropy of
mixing of clusters in an OCP.  The entropy of an ideal system of $N_i$
indistinguishable particles of mass $m_i$ at temperature $T$ and
within a volume $V$ is
\begin{equation}
\frac{S_i(V)}{\kB} = N_i \left [ \ln \left (\frac{V}{N_i} \right ) +
\frac{5}{2} + \frac{3}{2} \ln \left (\frac{2 \pi m_i \kB T}{h^2}
\right ) \right ].
\end{equation}
The ideal entropy of mixing is thus
\begin{equation}
\Smix = \infsum{1} \left [S_i (\Vtot) - S_i (V_i) \right ],
\end{equation}
where
\begin{equation}
V_i = \frac{Z_i N_i}{\nel}.
\end{equation}
is the fractional volume occupied by the $N_i$ clusters of charge
$Z_i$ and
\begin{equation}
\Vtot = \infsum{1} V_i = \frac{\Zavg \Ntot}{\nel},
\end{equation}
is the total volume of the system.  The $V_i$'s are set by demanding
both local charge neutrality and that $\nel$ is constant throughout
the medium.  It follows that
\begin{equation}
\frac{\Smix}{\kB} = \infsum{1} N_i \ln \left (\frac{\Vtot}{V_i} \right)
\end{equation}
and hence
\begin{equation}
\frac{\Smix}{\kB} = - \infsum{1} N_i \ln \left (\frac{Z_i N_i}{\Zavg
\Ntot} \right ).
\end{equation}

\end{appendix}

\bibliographystyle{revtex}

\end{document}